\documentclass[preprint]{aastex}

\usepackage{natbib}
\shorttitle{V2281~Cyg} 
\shortauthors{J.-R. KOO ET AL.} 

\begin{document}
\title{The $Kepler$ eclipsing binary V2281~Cygni with twin stars}
\author{Jae-Rim Koo$^{1,2}$, Jae Woo Lee$^{1,3}$, Kyeongsoo Hong$^{1,4}$}
\affil{$^{1}$Korea Astronomy and Space Science Institute, Daejeon (34055), Republic of Korea; \email{koojr@kasi.re.kr}
$^{2}$Dept. of Astronomy and Space Science, Chungnam National University, Daejeon (34134), Republic of Korea \\
$^{3}$Astronomy and Space Science Major, University of Science and Technology, Daejeon (34113), Republic of Korea \\
$^{4}$Institute for Astrophysics, Chungbuk National University, Cheongju (28644), Republic of Korea}

\begin{abstract}

We present the physical properties of the eclipsing binary V2281~Cyg which shows a light-time effect due to a supposed tertiary component from its eclipse timing variation according to the $Kepler$ observations. 
The high-resolution spectra and $BVR$ photometric data of the system were obtained at Bohyunsan Optical Astronomy Observatory and Mount Lemmon Optical Astronomy Observatory, respectively. 
To determine the fundamental parameters of the eclipsing pair and its circumbinary object, we simultaneously analyzed the radial velocities, light curves, and eclipse times including the $Kepler$ data.
The masses and radii for the primary and secondary stars were determined with accuracy levels of approximately 2\% and 1\%, respectively, as follows:
$M_{1} = 1.61 \pm 0.04$ M$_\odot$ and $M_{2} = 1.60 \pm 0.04$ M$_\odot$, $R_{1} = 1.94 \pm 0.02$ R$_\odot$ and $R_{2} = 1.93 \pm 0.02$ R$_\odot$.
If its orbit is coplanar with the eclipsing binary, the period and semi-major axis of the third body were calculated to be $P_{3b}=4.1$ years and $a_{3b}=4.06$ au, respectively, and its mass is $M_{3b}=0.75$ M$_\odot$.
The evolutionary state of the system was investigated by comparing the masses and radii with theoretical models. The results demonstrate that 
V2281~Cyg is a detached eclipsing binary which consists of twin main-sequence stars with an age of 1.5 Gyr.

\end{abstract}


\section{Introduction}

The $Kepler$ mission \citep{borucki2010} provides exquisitely precise and continuous photometric data unaffected by atmospheric effects.
Though the main goal of the mission is to search for habitable planets using the transit method, numerous pulsating stars and eclipsing binaries have also been observed during its mission. 
Many pulsating stars such as $\delta$~Sct and $\gamma$~Dor types in eclipsing binary systems have been investigated from the $Kepler$ light curves (LCs) \citep{southworth2011,lee2014,guo2016}. 
Moreover, orbital period studies based on long and continuous photometry provide the opportunity to detect additional objects orbiting binary systems. 
In total, 2878 eclipsing binaries have been observed in the $Kepler$ main field \citep{kirk2016}, and \citet{borkovits2016} identified 222 systems showing a light-travel-time effect, which may stem from the existence of a third body.

Binary systems serve as astrophysical laboratories because they allow us to deduce the properties of stars more accurately than we can with single stars. The fundamental parameters (mass and radius) of each component star can be determined precisely and directly using double-lined spectroscopic data and photometric time-series observations. 
Detached eclipsing binaries with accuracy levels within a few percent are very important objects as distance indicators, calibrators of other astrophysical scaling relationships, and tracers for binary evolutionary processes \citep{southworth2015}. Because the $Kepler$ LCs are highly precise, more accurate physical parameters can be determined from simultaneous analyses with other observational data. 
Many $Kepler$ eclipsing binaries have been studied together with spectroscopic observations to determine their fundamental properties and evolutionary statuses.

V2281~Cyg (RA=$19\fh25\fm06\fs90$, Dec=$+45\degr56\arcmin03\arcsec$; KIC~9402652; $K_p$=11.82; $V$=12.11; $B-V$=+0.44) was mentioned 
as an Algol-type eclipsing binary with an orbital period of $P=1.073$ days by \citet{diethelm2001}. 
The object is considered an eclipsing binary candidate containing a third component in the $Kepler$ data. 
The times of minimum light of the system showed four-year sinusoidal variations and a hierarchical third object was expected 
\citep{gies2015,zasche2015,borkovits2016}. In the $Kepler$ LC, the primary and secondary eclipse depths are very similar, indicating that 
the two components have nearly identical temperatures and luminosities \citep{zasche2015}.

Previous studies of V2281~Cyg were conducted with only single-band photometric data from the $Kepler$ mission and concentrated on period variations. 
In this paper, we present accurate physical parameters from simultaneous analyses of our multi-band LCs and radial velocities (RVs) with the $Kepler$ LC and timing data.
Section 2 describes $BVRK_p$ photometric and spectroscopic observations, and the spectroscopic analyses are presented in section 3. 
The binary synthesis and its absolute parameters are provided in section 4.
The summary and discussions are given in section 5.

\section{Observations and Reductions}

\subsection{$Kepler$ Photometry}

\begin{figure}
\includegraphics[trim=20 155 100 120, clip, width=\columnwidth]{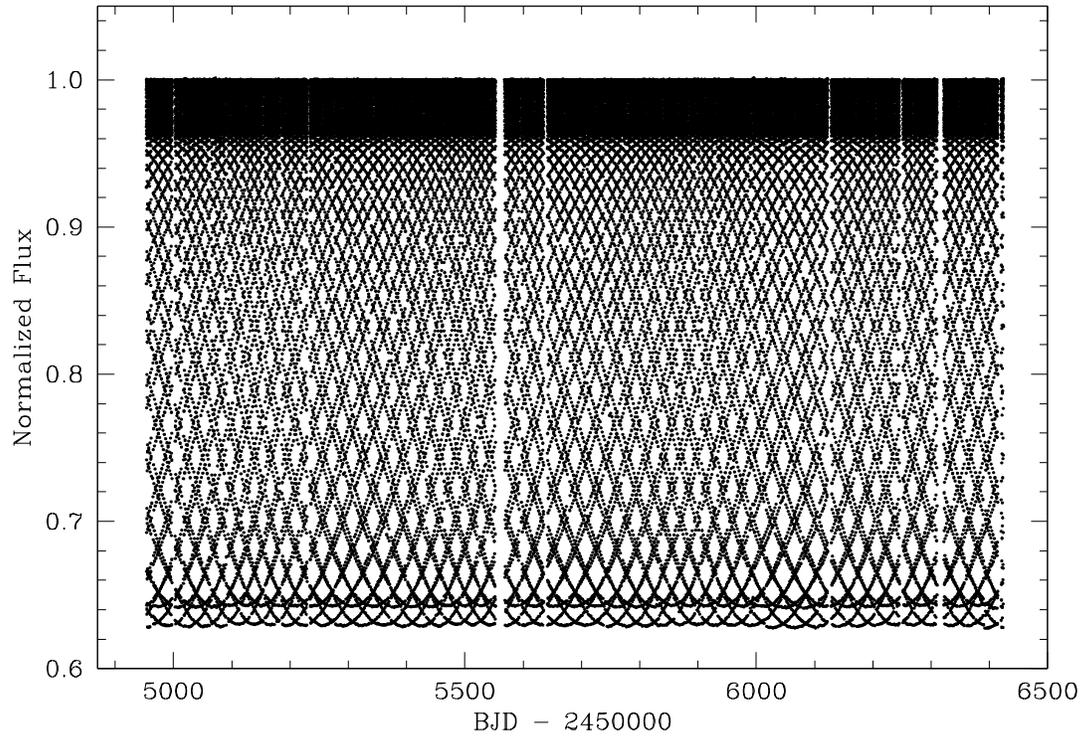}
\caption{Detrended and normalized $Kepler$ light curve of V2281~Cyg from Q0 to Q17. }
\label{Fig1}
\end{figure}

V2281~Cyg was observed by the $Kepler$ satellite from 2009 to 2013. Long-cadence observations of 29.42-min sampling were obtained during Quarters 0 to 17. 
The contamination level of the observations is estimated to be 0.012, where a value of 0 implies no contamination and 1 implies all background. 
We used the Simple Aperture Photometry (SAP) output by the $Kepler$ data processing pipeline\footnote{\url{http://archive.stsci.edu/kepler/}}. 
Since the entire data set has many systematic jumps and trends, we split the whole LC into 50 subsets based on the jumps and applied second or third order polynomial fits to the out-of-eclipse envelope ( 0.15 $<$ phase $<$ 0.35 and 0.65 $<$ phase $<$ 0.85 ) of each segment. 
\textbf{Some points that deviated by more than 5-sigma from light curve pattern were removed from the LC. }
Finally, the fluxes were normalized to unity and the de-trended LC is displayed in Figure~\ref{Fig1}.

\subsection{Ground-based Photometry and Spectroscopy}

New photometric observations for V2281~Cyg were performed using an ARC 4k CCD camera attached to a 1.0-m telescope 
at LOAO (Mount Lemmon Optical Astronomy Observatory) in Arizona, USA.
\cite{lee2012} had described the information about the detector and the telescope. 
1015 $V$-band observations to measure the primary eclipsing times were performed in two nights in 2014. 
A total of 1961 sets of $BVR$ images were observed during 8 nights in 2015. 
We did not observed standard stars because there are many stars in the observing field. 
The observation logs are listed in Table~\ref{tab0}. 
Bias subtraction and flat field were applied to the observed images for each night.

Because the light variations of the binary star could be affected by the presence of a neighboring star with separation of about 10 arcsec on the east-southern side,
we performed a PSF-fitting photometry on the observed images using the IRAF/DAOPHOT package. 
GSC~03543-01496 ($K_{\rm{p}}$=12.94; $V$=13.03; $B-V$=+0.79) was selected as a comparison star, with no brightness variation was detected against measurements of the nearby monitoring stars. 
Because we did not observed any standard field for photometric calibration, we tried to standardize all stars in the observed field using the photometric catalogue APASS DR9 \citep{henden2016}. 
Since APASS catalogue has no $R$ magnitudes, we transformed the SDSS system to standard $R$ magnitudes using the equation of \citet{jester2005}. 
Then, an ensemble normalization \citep{gilliland1988} was applied to obtain the standard magnitudes for our $BVR$ photometric data. 
This technique can correct the effects of a color term and x, y positions. We used about 100 stars for the calibration standards. 
Table~\ref{tab1} lists the standard $V$ and color indices of ($B-V$) and ($V-R$) for four characteristic phases. 

\begin{figure}
\includegraphics[width=\columnwidth]{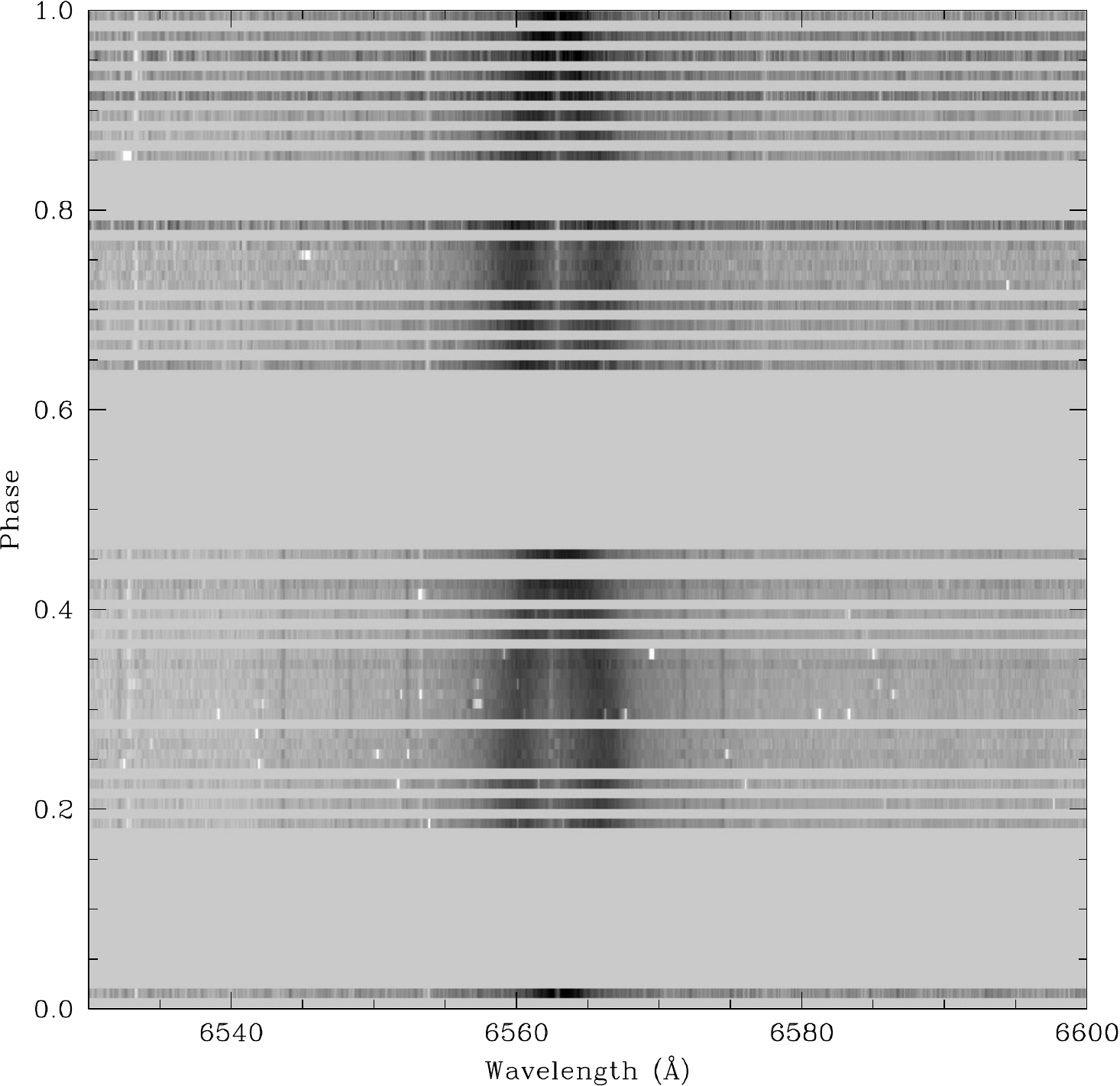}
\caption{Trailed spectra of H$_\alpha$ region for V2281~Cyg. The absorption lines of similar strength from two components can be identified easily, shifted through orbital phases.}\clearpage

\label{Fig2}
\end{figure}

Spectroscopic observations were carried out to measure the RVs and the effective temperatures of both components.
A total of 40 high-resolution spectra of V2281~Cyg were obtained in 2015 
using the fiber-fed spectrograph BOES \cite[Bohyunsan Optical Echelle Spectrograph;][]{kim2007} 
attached to the 1.8-m reflector at BOAO (Bohyunsan Optical Astronomy Observatory) in Korea. 
The BOES spectra covered a wavelength region from 3500 to 10500 \AA. 
Because our program target is so faint to obtain spectra with high signal-to-noise ratio (SNR), 
the $2\times2$ binning mode and the largest fiber with a diameter of 300\,$\micron$ having a resolving power of R = 30,000 were selected. 
The exposure time was set to be 30 minutes, which corresponds to 2 per cent of the orbital period to prevent orbital smearing. 
\textbf{ThAr and THL images were also obtained for line identification and blaze function calibration every night.} 
Cosmic ray removal by \citet{pych2004} was applied to the observed images. 
Aperture extraction and flat fielding were utilized with the IRAF/ECHELLE package, and the SNR of the spectra at the region of $\sim$5500\AA~ was around 25.
Figure~\ref{Fig2} shows the trailed spectra of the H$_\alpha$ region for V2281~Cyg, where one can identify the orbital motions of both components easily.

\section{SPECTRAL ANALYSIS}

The RVs of each component for V2281~Cyg were obtained from the cross-correlation functions (CCFs) calculated with the xcsao task of the RVSAO package in the IRAF. 
During this process, we used a template spectrum, a synthetic model of $T_{\rm{eff}} = 6300$ K, $v$sin$i = 90$\,km\,s$^{-1}$, log $g$ = 4.0, and [M/H] = 0.5 as obtained in this and discussion sections. 
The CCFs were applied to the \ion{Mg}{1} Triplet (5167, 5173 and 5184 \AA) region of each observed spectrum. 
Because the observed spectra have signals from primary and secondary stars, there were two peaks in the CCFs. 
We performed the Gaussian fits on the peaks with MPFIT \citep{markwardt2009}. 
The RV errors were calculated using the equation of \citet{kurtz1998}. 
Only 34 RVs were determined for both components because remained 6 spectra were too blended to separate the RVs. 
The RVs and their errors are listed in Table~\ref{tab2}. 

To obtain each spectra of the primary and secondary stars for determination of the effective temperatures and rotational velocities, 
we attempted to disentangle the observed spectrum using the FDB\textsc{inary}\footnote{\url{http://sail.zpf.fer.hr/fdbinary/}} code of \citet{ilijic2004}. 
We selected 14 spectra of hydrogen lines (H$_\alpha$, H$_\beta$, H$_\gamma$, and H$_\delta$) and \ion{Mg}{2} 4481 \AA~absorption line regions with relatively high SNR of different phases except near the eclipses. 
In the runs, the orbital period and RV amplitudes of both components were fixed to be $P=1.073167$ days, $K_1=151$\,km\,s$^{-1}$, and $K_2=152$\,km\,s$^{-1}$ based on the binary modeling results in the following section. 
The light factors were set to 0.5 because the depths of primary and secondary minima were almost same. 

\citet{slettebak1975} measured the rotational velocities of A-F type stars using the full-widths at the half intensity (FWHM) of the \ion{Mg}{2} 4481\AA~absorption line. We measured the FWHMs of the lines for both stars, and the rotational velocities were determined to be $v_{1}$sin$i = 80 \pm 10$\,km\,s$^{-1}$ and $v_{2}$sin$i = 85 \pm 10$\,km\,s$^{-1}$, respectively, from the linear fit and the scatters of the measurements by \citet{slettebak1975}. 

\begin{figure}
\includegraphics[trim=0 50 60 140, clip=true, width=\columnwidth]{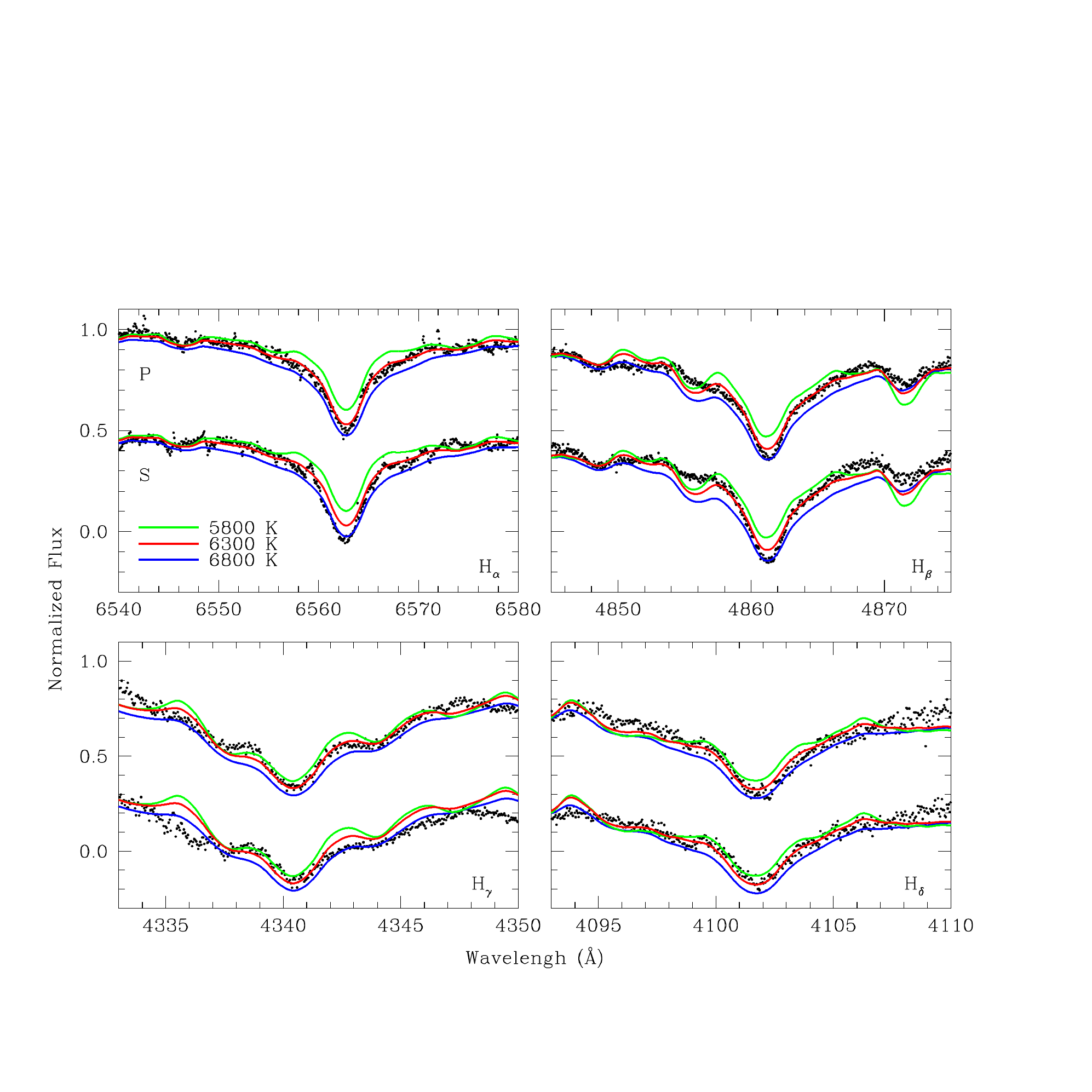}
\caption{Disentangled and synthetic spectra of hydrogen line regions (H$_\alpha$, H$_\beta$, H$_\gamma$, and H$_\delta$) for the primary and secondary components ($-0.5$ shifted). Dots display the observations and solid lines represent the synthetic models of $T_{\rm{eff}} = 5800,~6300,$ and 6800 K, respectively. }
\label{Fig3}
\end{figure}

For measuring the effective temperatures of the two components, we applied iSpec \citep[the integrated spectroscopic framework;][]{blanco2014} to the disentangled spectra of four hydrogen line regions. In this process, we adopted MARCS model atmospheres \citep{gustafsson2008} and rotational velocity of 90\,km\,s$^{-1}$. The metallicity was assumed as [M/H] = 0.5 deduced in the discussion section via evolutionary model fitting. The effective temperatures were calculated from minimizing the difference between observed and synthetic spectra. 
Averaged effective temperatures with standard deviations of the primary and secondary stars were determined to be $T_{\rm{eff},1} = 6355 \pm 180$ K and $T_{\rm{eff},2} = 6265 \pm 200$ K, respectively. 
Figure~\ref{Fig3} displays the disentangled hydrogen absorption lines of both components with the synthetic spectra of $T_{\rm{eff}}$ = 5800, 6300, and 6800 K.

\section{Binary Modeling and Absolute Dimensions}

\begin{figure}
\includegraphics[trim=0 30 120 80, clip=true, width=\columnwidth]{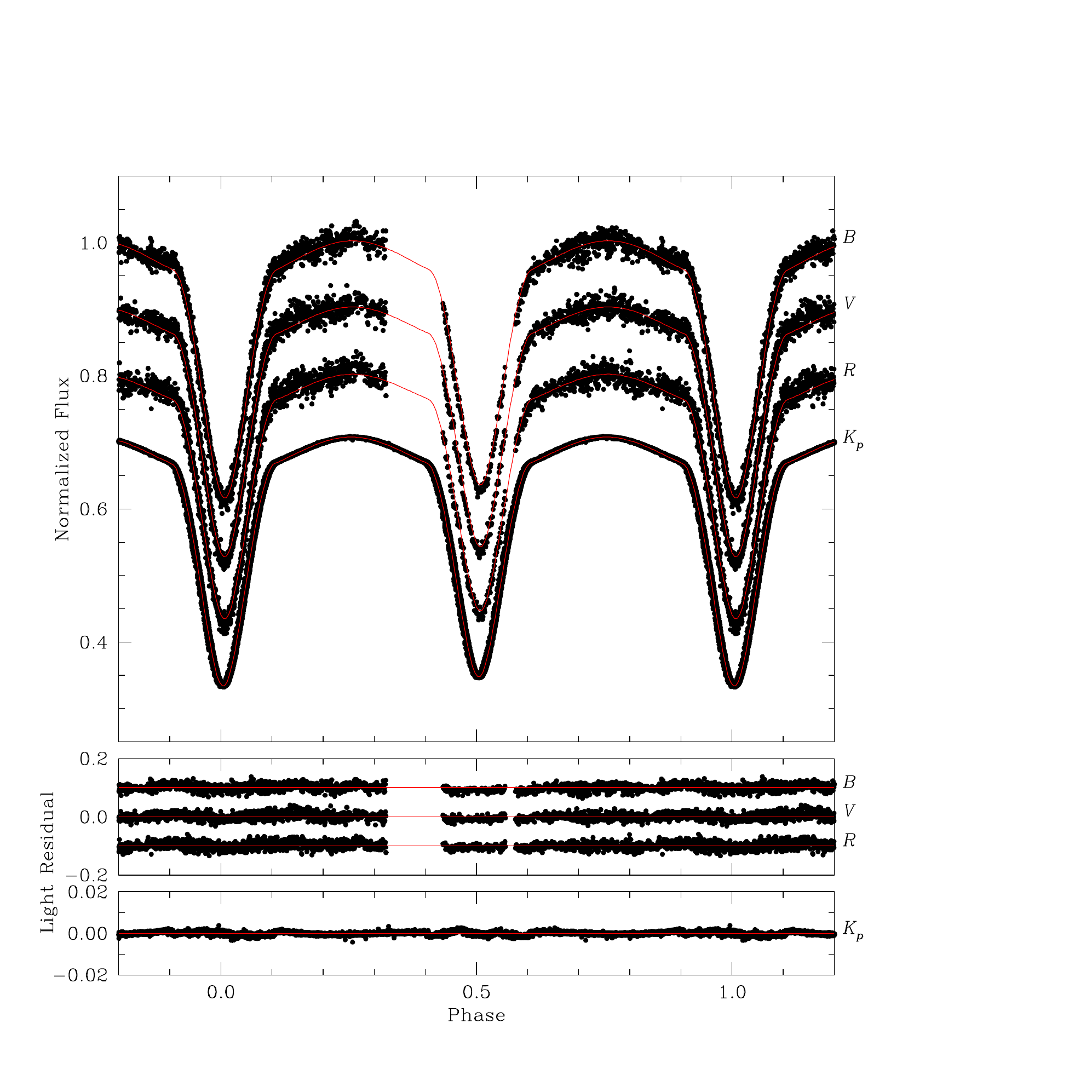}
\caption{$BVR$ and $K_p$ light curves of V2281~Cyg with fitted models. Circles are individual observations from LOAO and the $Kepler$ satellite. Lower panels display the differences between measurements and models.}
\label{Fig4}
\end{figure}

To determine the binary parameters of V2281~Cyg, our double-lined RVs, $BVR$ and $Kepler$ LCs, and the eclipse timings were analyzed using the 2015 version of the Wilson-Devinney synthesis code \citep[][hereafter W-D]{wilson1971,wilson2014}. The times of all observations were transformed from JD based on UTC into TDB-based BJD using the online applets\footnote{\url{http://astroutils.astronomy.ohio-state.edu/time/}} by \citet{eastman2010}. 
As shown in Figure~\ref{Fig4}, the depths of the primary and secondary minima are very similar to each other, which indicates that the two components have nearly identical temperatures and luminosities. 
To examine the period variation due to the third body suggested by previous investigators, we obtained seven times of minimum light from our photometric data by applying the method of \citet{kwee1956} which can determine the time and error in the symmetric light curve. 
Minimum time and their errors for $BVR$ filters were calculated by applying the error-weighted mean and errors, 
and they are listed in Table~\ref{tab3}. In the Table, Min I and II indicate primary and secondary minima, respectively.  
A total of 2626 minimum times for this system were compiled by \citet{zasche2015} from the previous observations including the $Kepler$ and their own data. 
However, we selected only 2543 minimum times from the $Kepler$ observations to ours for analyses because the earlier data contains a large scatter.

\begin{figure}
\includegraphics[trim=0 20 120 200, clip=true, width=\columnwidth]{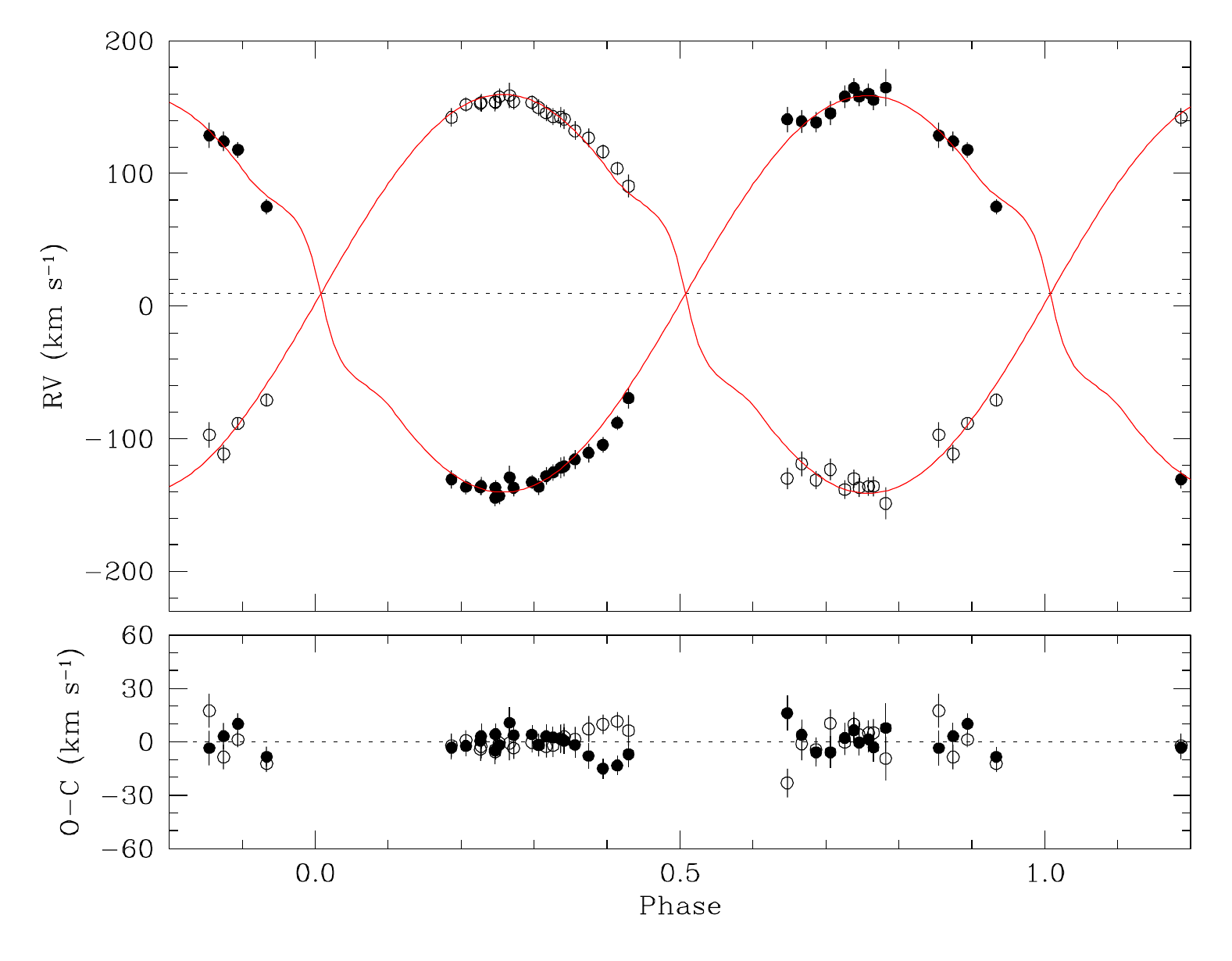}
\caption{RV curves of V2281~Cyg with fitted models. Filled and open circles represent the RV measurements for the primary and secondary components, respectively. The dotted line refers to the system velocity of $9.6$\,km\,s$^{-1}$ in the upper panel. Residuals between observations and models are displayed in the lower panel.}
\label{Fig5}
\end{figure}

We first adopted mode 2 of the W-D synthesis code because the binary was reported as a detached system by \citet{zasche2015}. 
Various modes were applied to check the possibility of the other Roche configurations but only the detached model was acceptable. 
In the procedure, the orbital eccentricity of the binary was remained zero within its error. 
The third-light was not considered because the calculated values were not physically meaningful in the run. 
The input effective temperature for the primary star was fixed to be $T_{\rm{eff},1} = 6355$ K from 
our spectral analyses. 
The gravity-darkening exponents and bolometric albedoes were assumed to 
be $g_{1,2}$=0.32 and $A_{1,2}$=0.5, respectively, which is suitable for stars with convective envelopes. 
The logarithmic bolometric ($X$, $Y$) and monochromatic ($x$,$y$) limb-darkening coefficients were calculated from the values of \citet{van1993}. 
The inclination of the third body ($i_{3b}$) was set to be identical to that of the binary ($i$) because they would have formed in the same gas cloud. 
The W-D syntheses were repeated until the correction for each parameter become smaller than its 
standard deviation computed from the differential correction (DC) program of the W-D. 
The final solutions are listed in Table~\ref{tab4} and the values with parenthesized errors signify the adjusted parameters. 
The primary and secondary stars were expressed with the subscripts 1 and 2, respectively.
Our result indicates that the eclipsing binary consists of twin stars. 
The mass ratio was calculated to be $q = 0.994$, and the radii and effective temperatures of both components are almost same each other. 
Figs.~\ref{Fig4} and \ref{Fig5} display the light and RV curves with fitted models, respectively. 

\begin{figure}
\includegraphics[trim=0 50 120 200, clip=true, width=\columnwidth]{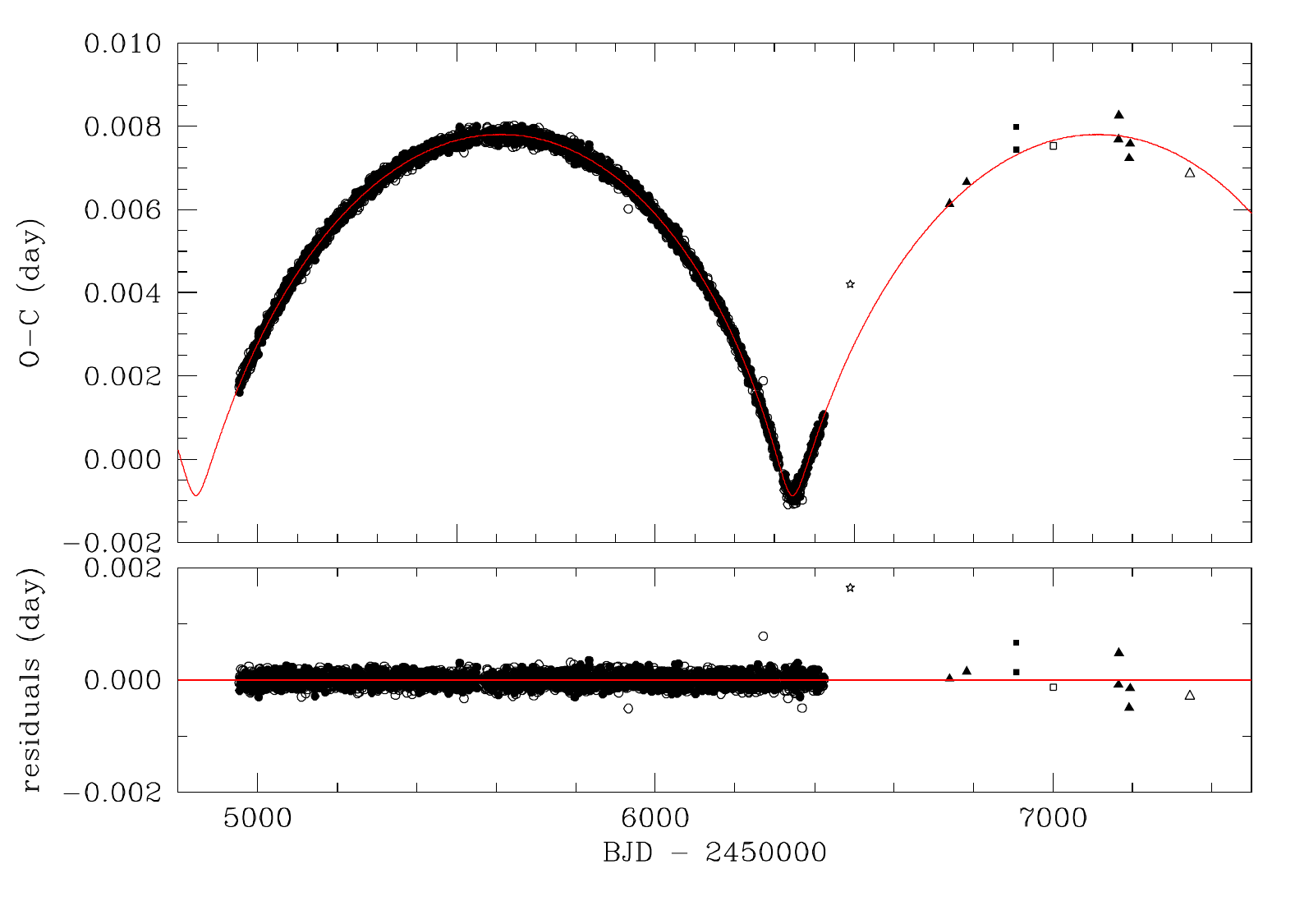}
\caption{$O-C$ diagram for eclipse timing of V2281~Cyg with fitted model. Filled and open symbols represent the primary and secondary minima, respectively. 
Circles, squares, and a star symbol are data from $Kepler$ mission, \citet{zasche2015}, and \citet{diethelm2014}, respectively. Data of this study are represented with triangles. Lower panel displays the residuals between measurements and our model.}
\label{Fig6}
\end{figure}

Additionally, the third companion orbiting the eclipsing binary has a mass of $m_3=0.75$ M$_\odot$ as calculated from the value of $M_3$/($M_1$+$M_2$) = 0.2319, a period of 4.1 years, and a semi-major axis of 4.06 au.
The tertiary mass is in good agreement with the mass from the mass function $f(m_{3})\sim0.026$ for the third-body in previous eclipse timing studies \citep{gies2015,zasche2015,borkovits2016}.
If the tertiary object is the main-sequence star, the object is a K3 type having $T_{\rm{eff},3}\approx4700$ K, and its contribution to the total light is calculated to be $l_3\sim2.5 \%$. However, the third-light rates calculated with W-D synthesis code were not physically meaningful, and no spectroscopic features were detected in our spectra. 
The eclipse timing $O-C$ diagram of V2281~Cyg is displayed in Figure~\ref{Fig6} with our fit.

From our consistent light and RV solutions, the absolute dimensions for each component were calculated using the JKTABSDIM code 
\citep{southworth2005b} and listed in Table~\ref{tab5}. The luminosity ($L$) and bolometric magnitude (M$_{\rm{bol}}$) were computed by adopting $T_{\rm eff}$$_\odot$ = 5780 K and $M_{\rm bol}$$_\odot$ = +4.73 for solar values. 
The bolometric corrections (BCs) were calculated from the relation between $\log T_{\rm eff}$ and BC given by \citet{torres2010}. 
The synchronized rotational velocities of both components were calculated to be $v_{1,\rm{sync}}$sin$i$ = 90 $\pm$ 1\,km\,s$^{-1}$ and $v_{2,\rm{sync}}$sin$i$ = 90 $\pm$ 1\,km\,s$^{-1}$, respectively. These were comparable to the measured rotational velocities within their errors in the previous section.

For distance determination to the system, we adopted interstellar reddening of $E(B-V) = 0.08$ as derived from the difference between the intrinsic $B-V$ color for the temperature \citep{flower1996} and observed one. 
Then the distance was calculated to be 750$\pm$50 pc, in good agreement with 781$\pm$215 pc as calculated with the trigonometric parallax ($1.28 \pm 0.33$ mas) from the $Gaia$ DR1 \citep{gaia2016}.

\section{Summary and Discussions}
In this paper, we present the physical properties of the eclipsing binary V2281~Cyg from simultaneous analyses of new spectroscopic and multi-band photometric observations with the $Kepler$ LC and eclipse timings. 
The effective temperatures of both component stars were determined to be $T_{1} = 6355$ K and $T_{2} = 6284$ K from our analyses.
The RVs for each star were measured and the absolute parameters were derived from W-D syntheses of $BVRK_p$ LCs, RVs, and the minimum times. The masses and radii of the inner components were calculated to be $M_{1} = 1.61 \pm 0.04$ M$_\odot$ and $M_{2} = 1.60 \pm 0.04$ M$_\odot$, $R_{1} = 1.94 \pm 0.02$ R$_\odot$ and $R_{2} = 1.93 \pm 0.02$ R$_\odot$, respectively.  

\begin{figure}
\includegraphics[trim=0 0 120 150, clip=true, width=\columnwidth]{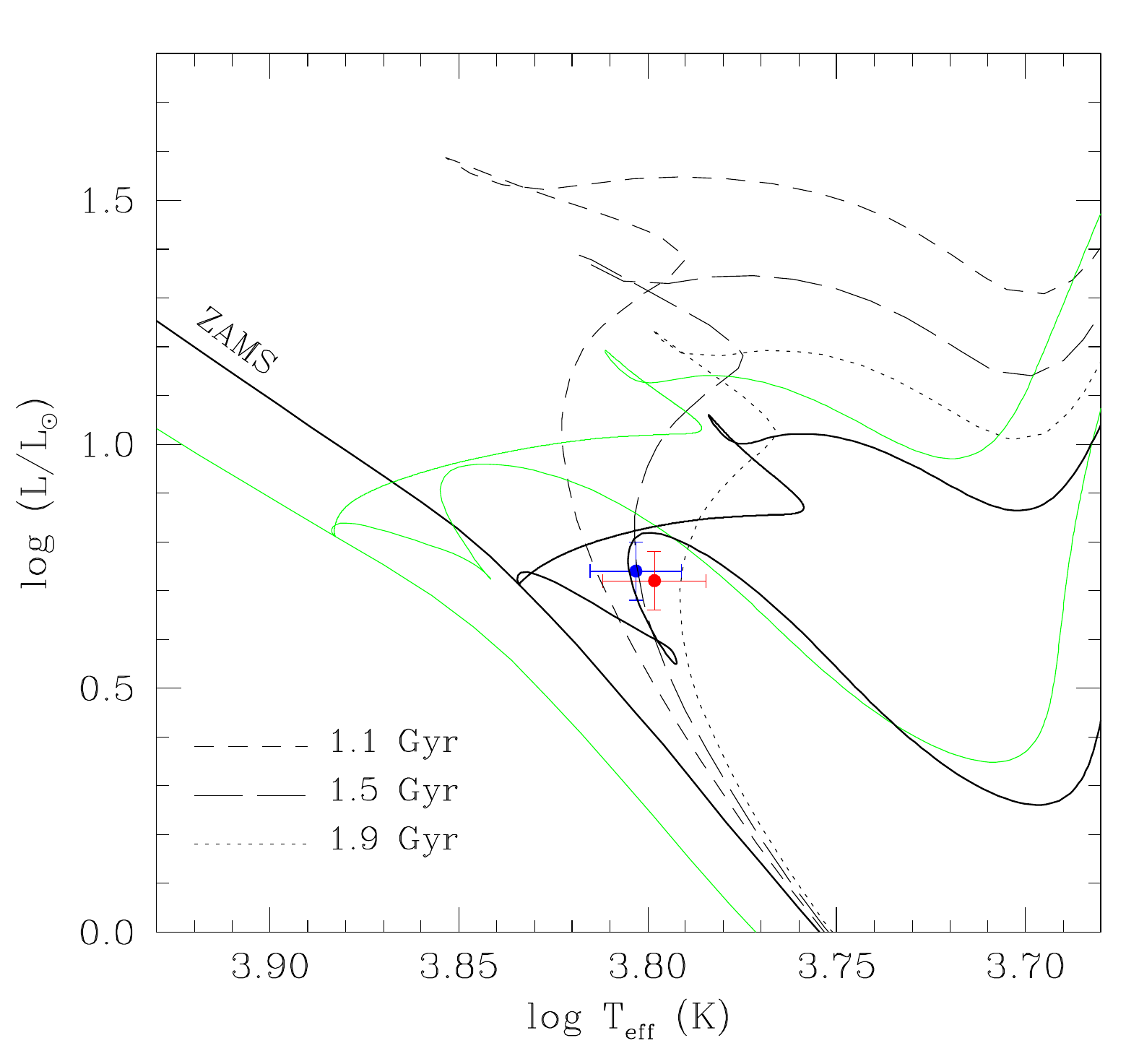}
\caption{The positions on the HR diagram of both components of V2281~Cyg and comparison with PARSEC evolutionary models \citep{bressan2012}. 
The primary and secondary components were displayed in blue and red dots with their errors, respectively. 
The black and green solid lines represent the evolutionary tracks of $Z = 0.06$ and $Z = 0.017$ (solar value) for $M$ = 1.6 M$_{\odot}$, respectively. Three isochrones of age = 1.1, 1.5, and 1.9 Gyrs were displayed with different lines.}
\label{Fig7}
\end{figure}

From our binary modeling results, the fill-out-factors ($f$ = $\Omega_{\rm{in}}$/$\Omega$ ) of both component stars indicate that V2281~Cyg is a detached eclipsing binary system which has no mass transfer between the stars. Therefore, we can estimate the evolutionary status and age of the system by comparing our accurate absolute properties with stellar evolutionary models.
Using PARSEC evolutionary tracks \citep{bressan2012}, we tried to find the model satisfying our results. 
For the solar abundant evolutionary track of $M$ = 1.6 M$_\odot$, both components of V2281~Cyg are located near the pre-main-sequence (PMS) stage, as shown in Figure~\ref{Fig7}. 
However, the binary system could not be in the PMS stage because any emission features in the hydrogen lines were absent in our spectra and there is no star-forming region within 5 degrees. 
Thus, we could conclude that V2281~Cyg consists of two 1.6 M$_\odot$ main-sequence stars of an age\,$\approx$\,1.5 Gyr and $Z$ = 0.06. 
In Figure~\ref{Fig7}, the black solid line represents the evolutionary track of $M$ = 1.6 M$_\odot$ and $Z$ = 0.06, and three isochrones of age = 1.1, 1.5, and 1.9 Gyr are plotted with dashed lines. 
However, it seems that the higher abundant models are required for better agreement between models and observations.

\acknowledgements{
This work was supported by KASI (Korea Astronomy and Space Science Institute) grant 2017-1-830-03. 
J.-R. Koo was supported by Basic Science Research Program through the National Research Foundation of Korea (NRF) funded by the Ministry of Education (grant 2017R1A6A3A01002871).
One of the authors, K. Hong was supported by the grant 2017R1A4A1015178 of National Research Foundation of Korea. 
We have used the Simbad database maintained at CDS, Strasbourg, France.
This paper includes data collected by the $Kepler$ mission. 
Funding for the Kepler mission is provided by the NASA Science Mission directorate.
This work has made use of data from the European Space Agency (ESA)
mission {\it Gaia} (\url{http://www.cosmos.esa.int/gaia}), processed by
the {\it Gaia} Data Processing and Analysis Consortium (DPAC,
\url{http://www.cosmos.esa.int/web/gaia/dpac/consortium}). Funding
for the DPAC has been provided by national institutions, in particular
the institutions participating in the {\it Gaia} Multilateral Agreement.
}

\newpage

\begin{table}
\caption{Photometric observation log of V2281~Cyg}
\label{tab0}
\begin{tabular}{ccrc} 
\hline 
  Date     &    Filter     &  $N_{obs}$  & Weather        \\  
\hline                                                               
2014.03.22 &      $V$      &    420      &  Clear         \\        
2014.05.04 &      $V$      &    595      &  Clear         \\        
2015.05.21 &    $BVR$      &    573      &  Partly cloudy \\        
2015.05.22 &    $BVR$      &    594      &  Partly cloudy \\        
2015.05.25 &    $BVR$      &    858      &  Partly cloudy \\        
2015.06.16 &    $BVR$      &   1032      &  Clear         \\        
2015.06.17 &    $BVR$      &    960      &  Clear         \\        
2015.06.19 &    $BVR$      &    621      &  Partly cloudy \\        
2015.06.21 &    $BVR$      &    846      &  Clear         \\        
2015.11.17 &    $BVR$      &    399      &  Clear         \\        
\hline
\end{tabular}
\end{table}

\begin{table}
\caption{Standard magnitudes and color indices at four characteristic phases of V2281~Cyg}
\label{tab1}
\begin{tabular}{cccc} 
\hline
 phase      &          $V$     &    $B-V$          &   $V-R$          \\  
\hline                                                                
0.00        & 12.27 $\pm$ 0.02 &  0.56 $\pm$ 0.03  &  0.25 $\pm$ 0.03  \\        
0.25        & 11.78 $\pm$ 0.01 &  0.54 $\pm$ 0.02  &  0.26 $\pm$ 0.01 \\        
0.50        & 12.23 $\pm$ 0.01 &  0.54 $\pm$ 0.01  &  0.26 $\pm$ 0.01 \\        
0.75        & 11.78 $\pm$ 0.01 &  0.53 $\pm$ 0.02  &  0.25 $\pm$ 0.02 \\        
\hline
\end{tabular}
\end{table}

\begin{table}
\caption{Radial velocities for V2281~Cyg}
\label{tab2}
\begin{tabular}{lrrrr} 
\hline
       BJD   &  V$_1$   &   $\sigma_1$               &  V$_2$   &  $\sigma_2$      \\
 {$-$2450000}  & \multicolumn{2}{c}{(km\,s$^{-1}$)}    & \multicolumn{2}{c}{(km\,s$^{-1}$)}    \\
\hline
 7123.2736 &  $-$136.6 &    6.2 &     152.9 &   5.8    \\ 
 7123.2952 &  $-$144.6 &    6.9 &     153.9 &   6.5    \\ 
 7123.3164 &  $-$129.1 &    9.5 &     158.9 &   8.9    \\ 
 7130.2615 &     164.4 &    7.0 &  $-$130.1 &   7.2    \\ 
 7130.2830 &     160.4 &    6.9 &  $-$135.9 &   7.4    \\ 
 7156.1409 &     128.8 &    9.5 &   $-$97.0 &   9.6    \\ 
 7156.1621 &     124.1 &    6.9 &  $-$111.4 &   7.4    \\ 
 7156.1832 &     117.9 &    4.2 &   $-$88.4 &   5.9    \\ 
 7156.2255 &      75.0 &    4.9 &   $-$70.8 &   5.7    \\ 
 7159.1375 &     140.9 &    8.0 &  $-$129.9 &   9.6    \\ 
 7159.1586 &     139.4 &    9.3 &  $-$118.8 &   8.6    \\ 
 7159.1798 &     138.7 &    6.9 &  $-$131.0 &   7.6    \\ 
 7159.2009 &     145.5 &    7.9 &  $-$123.2 &   9.0    \\ 
 7159.2220 &     158.1 &    7.3 &  $-$138.3 &   8.7    \\ 
 7159.2431 &     158.0 &    7.0 &  $-$136.9 &   7.4    \\ 
 7159.2643 &     155.5 &    7.7 &  $-$135.8 &   7.8    \\ 
 7159.2823 &     164.8 &   12.1 &  $-$148.8 &  13.9    \\ 
 7307.9336 &  $-$136.2 &    6.3 &     149.8 &   6.4    \\ 
 7307.9548 &  $-$125.5 &    6.4 &     142.9 &   6.1    \\ 
 7307.9711 &  $-$120.8 &    7.5 &     141.1 &   7.4    \\ 
 7308.0071 &  $-$110.8 &    7.3 &     127.0 &   7.0    \\ 
 7308.0282 &  $-$104.7 &    5.5 &     116.5 &   6.0    \\ 
 7308.0493 &   $-$88.1 &    5.3 &     103.8 &   5.4    \\ 
 7308.0658 &   $-$69.4 &    8.7 &      90.5 &   7.5    \\ 
 7308.9488 &  $-$143.2 &    6.1 &     158.0 &   6.1    \\ 
 7308.9699 &  $-$137.1 &    6.0 &     154.3 &   6.5    \\ 
 7308.9967 &  $-$132.8 &    5.6 &     153.6 &   5.3    \\ 
 7309.0177 &  $-$128.1 &    6.2 &     145.6 &   6.7    \\ 
 7309.0390 &  $-$122.0 &    7.9 &     142.3 &   7.8    \\ 
 7309.0602 &  $-$115.6 &    7.1 &     132.2 &   7.2    \\ 
 7309.9515 &  $-$130.6 &    6.7 &     142.4 &   6.8    \\ 
 7309.9727 &  $-$136.4 &    5.8 &     152.1 &   5.7    \\ 
 7309.9952 &  $-$135.7 &    6.8 &     153.5 &   6.9    \\ 
 7310.0163 &  $-$136.9 &    5.7 &     153.8 &   5.8    \\ 
\hline
\end{tabular}
\end{table}

\begin{table}
\caption{New minimum times for V2281~Cyg}
\label{tab3}
\begin{tabular}{lccc} 
\hline
        BJD ($-$2450000)   &    Error     & Min  & Filter \\
\hline
6739.93784 & $\pm$0.00008 &  I  & $V$\\
6782.86263 & $\pm$0.00005 &  I  & $V$\\
7164.88968 & $\pm$0.00007 &  I  & $BVR$\\
7165.96336 & $\pm$0.00014 &  I  & $BVR$\\
7191.71690 & $\pm$0.00011 &  I  & $BVR$\\
7193.86346 & $\pm$0.00011 &  I  & $BVR$\\
7344.63425 & $\pm$0.00013 &  II & $BVR$\\
\hline
\end{tabular}
\tablecomments{Min I and II indicate primary and secondary minima, respectively.
Minimum times and their errors for $BVR$ filters are error-weighted mean and errors.  }
\end{table}

\begin{table}
\caption{Simultaneous analysis results of V2281~Cyg}
\label{tab4}
\begin{tabular}{lcc}
\hline
       Parameter       &      Primary      &  Secondary             \\
\hline                                                            
$T_0$ (BJD)            &    \multicolumn{2}{c}{2457164.882006(11)}  \\
$P$ (days)             &    \multicolumn{2}{c}{1.073106806(9)}     \\
$a$ (R$_\odot$)        &    \multicolumn{2}{c}{6.513(5)}           \\
$q$                    &    \multicolumn{2}{c}{0.994(2)}           \\
$V_\gamma$ (km\,s$^{-1}$) &    \multicolumn{2}{c}{9.6(3.4)}            \\
$i$ (deg)              &    \multicolumn{2}{c}{79.866(4)}          \\
$T$ (K)                &     6,355(180)     &    6,284(200)        \\
$\Omega$               &      4.414(3)      &    4.399(4)         \\
$\Omega_{\rm{in}}$     &    \multicolumn{2}{c}{3.740(3)}          \\
$X$, $Y$               &      0.639, 0.236  &   0.640, 0.234       \\ 
$x_{B}$, $y_{B}$       &      0.811, 0.219  &   0.815,  0.212      \\
$x_{V}$, $y_{V}$       &      0.718, 0.271  &   0.723,  0.269      \\
$x_{R}$, $y_{R}$       &      0.625, 0.281  &   0.630,  0.279      \\
$x_{Kp}$, $y_{K_p}$    &      0.680, 0.271  &   0.684,  0.269      \\
$l$/($l_1+l_2$)$_B$    &      0.516(1)      &    0.484             \\
$l$/($l_1+l_2$)$_V$    &      0.512(1)      &    0.488             \\
$l$/($l_1+l_2$)$_R$    &      0.509(1)      &    0.491             \\
$l$/($l_1+l_2$)$_{K_p}$&      0.511(1)      &    0.489             \\
$r$ (pole)             &      0.2895(1)     &   0.2889(5)          \\
$r$ (point)            &      0.3152(2)     &   0.3149(8)          \\
$r$ (side)             &      0.2969(1)     &   0.2963(6)          \\
$r$ (back)             &      0.3078(2)     &   0.3073(7)          \\
$r$ (volume)$\rm ^a$   &      0.2981(2)     &   0.2975(7)          \\
 \\
                        \multicolumn{3}{l}{Third-body parameters}  \\
$a_{3b}$ (R$_\odot$)   &    \multicolumn{2}{c}{873.2(7)}           \\
$P_{3b}$ (days)        &    \multicolumn{2}{c}{1500.7(6)}          \\
$i_{3b}$ (deg)         &    \multicolumn{2}{c}{79.866}             \\
$e_{3b}$               &    \multicolumn{2}{c}{0.800(1)}           \\
$\omega_{3b}$ (rad)    &    \multicolumn{2}{c}{4.643(1)}           \\
$T_{03b}$ (BJD)        &    \multicolumn{2}{c}{2455682(2)}         \\
$M_{3}$/($M_{1}+M_{2}$) &    \multicolumn{2}{c}{0.2319}            \\
\hline
\multicolumn{3}{l}{$^a$ Mean volume radius.} \\
\end{tabular}
\end{table}

\begin{table}
\caption{Physical properties for V2281~Cyg}
\label{tab5}
\begin{tabular}{lcc}
\hline
      Parameter        &       Primary   &     Secondary    \\
\hline
Mass (M$_\odot$)       &       1.61 $\pm$ 0.04     &    1.60 $\pm$ 0.04     \\
Radius (R$_\odot$)     &       1.94 $\pm$ 0.02     &    1.93 $\pm$ 0.02     \\
log $g$ (cgs)          &       4.07 $\pm$ 0.01     &    4.07 $\pm$ 0.01     \\
log $L$ (L$_\odot$)    &       0.74 $\pm$ 0.06     &    0.72 $\pm$ 0.06     \\
M$_{\rm{bol}}$ (mag)   &       2.88 $\pm$ 0.12     &    2.92 $\pm$ 0.14      \\
BC (mag)               &         $-$0.01         &      $-$0.01         \\
$M_V$ (mag)            &       2.89 $\pm$ 0.12     &   2.93 $\pm$ 0.14      \\
Distance (pc)          &        \multicolumn{2}{c}{750 $\pm$ 50}   \\

\hline
\end{tabular}
\end{table}

\clearpage

\end{document}